\documentclass[a4paper,fleqn,usenatbib]{mnras}
\usepackage{graphicx}
\usepackage{amssymb}
\usepackage{amsmath}
\usepackage[T1]{fontenc}
\usepackage{ae,aecompl}
\usepackage{natbib,times}
\usepackage{lscape}
\usepackage{array}
\usepackage{setspace}
\usepackage{helvet}
\usepackage{float}
\newcommand{\PreserveBackslash}[1]{\let\temp=\\#1\let\\=\temp}
\newcolumntype{C}[1]{>{\PreserveBackslash\centering}p{#1}}
\newcolumntype{R}[1]{>{\PreserveBackslash\raggedleft}p{#1}}
\newcolumntype{L}[1]{>{\PreserveBackslash\raggedright}p{#1}}

\newcommand{\mc}{\multicolumn}

\begin{document}

\title[Dynamical state of galaxy clusters] 
{Dynamical state of galaxy clusters evaluated from X-ray images}

\author[Yuan, Han \& Wen]
       {Z. S. Yuan$^{1,2}$, 
        J. L. Han$^{1,2,3}$ \thanks{E-mail: hjl@nao.cas.cn} 
    and Z. L. Wen$^{1,2}$
\\
1. National Astronomical Observatories, Chinese Academy of Sciences, 
20A Datun Road, Chaoyang District, Beijing 100101, China\\
2. CAS Key Laboratory of FAST, NAOC, Chinese Academy of Sciences,
           Beijing 100101, China \\
3. School of Astronomy, University of Chinese Academy of Sciences,
           Beijing 100049, China 
}

\date{Accepted XXX. Received YYY; in original form ZZZ}

\label{firstpage}
\pagerange{\pageref{firstpage}--\pageref{lastpage}}
\maketitle


\begin{abstract}
X-ray images of galaxy clusters often show disturbed structures that
are indication of cluster mergers. As a complementary to our previous
work on dynamical state of 964 clusters observed by the {\it Chandra},
we process the X-ray images for 1308 clusters from {\it XMM-Newton}
archival data, together with the images of 22 clusters newly released
by the {\it Chandra}, and evaluate their dynamical state from these
X-ray images. The concentration index $c$, the centroid shift
$\omega$, the power ratio $P_3/P_0$ are calculated in circular regions
with a certain radius of 500 kpc, and the morphology index $\delta$ is
estimated within elliptical regions adaptive to the cluster size and
shape. In addition, the dynamical parameters for 42 clusters
previously estimated from {\it Chandra} images are upgraded based on
the newly available redshifts. Good consistence is found between
dynamical parameters derived from {\it XMM-Newton} and {\it Chandra}
images for the overlapped sample of clusters in the two datasets. The
dependence of mass scaling relations on dynamical state is shown by
using data of 388 clusters.
All data and related software are available on web-page
http://zmtt.bao.ac.cn/galaxy\_clusters/dyXimages/.
\end{abstract}

\begin{keywords}
  galaxies: clusters: general --- galaxies: clusters: intracluster
  medium --- galaxies: groups: general --- X-rays:galaxies:clusters
\end{keywords} 


\section{Introduction}

Clusters of galaxies are in various dynamical states. Some of them are
dynamically relaxed, but a large fraction of clusters are not in
dynamical equilibrium \citep[e.g.,][]{cb17,ltl+18,yh20}. The dynamical
state is a fundamental characteristic of galaxy clusters, in
additional to other properties such as the mass or the temperature of
intracluster medium (ICM).

The dynamical state of galaxy clusters can be quantitatively figured
out by various observational tracers. In optical, it can be roughly
indicated by the redshift distribution of member galaxies
\citep[e.g.,][]{yv77,ds88,wb90,ssg99,hmp+04,hph+09,rph18,yds+18},
which is the 1-dimensional (1D) information along the line of
sight. In principle, the dynamical state should be determined by the
3D velocities and mass distributions of all member galaxies and/or ICM
\citep[e.g.,][]{cd96,ets+10,evn+12}. However, it is very hard to get
the full 3D information in practice. The dynamical state can be well
represented with 2D maps, because if a cluster is disturbed in 2D map
it must be really disturbed in 3D though information on the radial
dimension is missing. Analyses of the 2D position distributions of
member galaxies have been tried for many years
\citep[e.g.,][]{gb82,wb90,fk06,rbp+07,as10,evn+12,wh13,ltl+18,
  kgm+19}. The largest sample of clusters is presented by
\citet{wh13}, who smoothed the optical flux-weighted maps of member
galaxies in the sky plane, and calculated the relaxation factor
$\Gamma$ for 2092 rich clusters. In X-ray, dynamical parameters of
clusters can be obtained by analyzing the images of hot intracluster
gas, though calculating the central cooling time
\citep[e.g.,][]{bfs+05,crb+07}, the concentration index
\citep[e.g.,][]{srt+08,ccb+15,zrs+17,lty+18}, the centroid shift
\citep[e.g.,][]{pfb+06,ceg+10,ceb+13,der+16}, the power ratio
\citep[e.g.,][]{bt95,wbsa13,wbc13}, or simply the morphology index
\citep{yh20}. In microwave band, the dynamical state of clusters can
also be derived from maps of Sunyaev-Zel'dovich (SZ) effect
\citep[e.g.,][]{cds+18}.

Previously, high quality X-ray images are difficult to get, therefore
dynamical parameters of only a limited sample (generally less than
100) of clusters were calculated
\citep[e.g.,][]{bt95,srt+08,ceg+10,ceb+13,wbsa13,der+16,zrs+17,lty+18}.
Recent years, the {\it Chandra} and the {\it XMM-Newton} X-ray
telescopes observed many clusters of galaxies. \citet{wbc13}
calculated the power ratios for 129 clusters in the redshift range of
$0.05<z<1.08$ from the {\it Chandra} and {\it XMM-Newton}
images. \citet{ajf+17} computed the concentration indexes for 214
clusters from {\it Chandra} images. \citet{zhk+20} derived the offset
of brightest cluster galaxies to the brightness peak of hot gas
indicated by SZ ({\it SPT}) or X-ray ({\it Chandra/XMM-Newton}) images
for 288 clusters. \citet{yh20} processed the released {\it Chandra}
images for 964 clusters, and calculated the concentration indexes, the
centroid shifts, the power ratios and also the morphology indexes.

The {\it XMM-Newton} X-ray telescope has observed more than 1000
clusters of galaxies with a larger field of view than the {\it
  Chandra}. As a continuation of \citet{yh20}, we calculate the
dynamical parameters for 1308 clusters in the archived {\it
  XMM-Newton} data. In Section~\ref{data}, we describe the procedures
for cluster finding from the data and image processing. In
Section~\ref{dp}, dynamical parameters are calculated, not only for
these 1308 clusters, but also for clusters newly released by the {\it
  Chandra}. In Section~\ref{appl}, the influence of dynamical state on
cluster mass estimation is discussed. Finally, a summary is given in
Section~\ref{summary}.

Throughout this paper, the flat $\Lambda$CDM cosmology is adopted with
$H_0=70$ km~s$^{-1}$ Mpc$^{-1}$, $\Omega_m=0.3$ and
$\Omega_{\Lambda}=0.7$.

\section{Cluster samples and image processing}
\label{data}

\subsection{Galaxy clusters in X-ray data archives}
\label{datacoll}

As in \citet{yh20}, we find clusters in the archived data by two
approaches: (1) targeted objects in proposed observations for galaxy
clusters; and (2) serendipitously detected clusters.

First, in the {\it XMM-Newton} Science Archive Search
system\footnote{http://nxsa.esac.esa.int/nxsa-web/\#search}, we select
the proposal category of ``Groups of galaxies, Clusters of galaxies
and Superclusters'', and find more than 2000 observations. For a
cluster with multiple observations, we choose the one with the longest
exposure time. For low redshift clusters (e.g., $z<0.05$), the field
coverage should also be cared. After checking the raw images,
discarding multiple observations and rejecting poor quality data, we
get the {\it targeted sample} of 892 clusters.

\begin{figure}
\centering
\includegraphics[angle=270,width=0.4\textwidth]{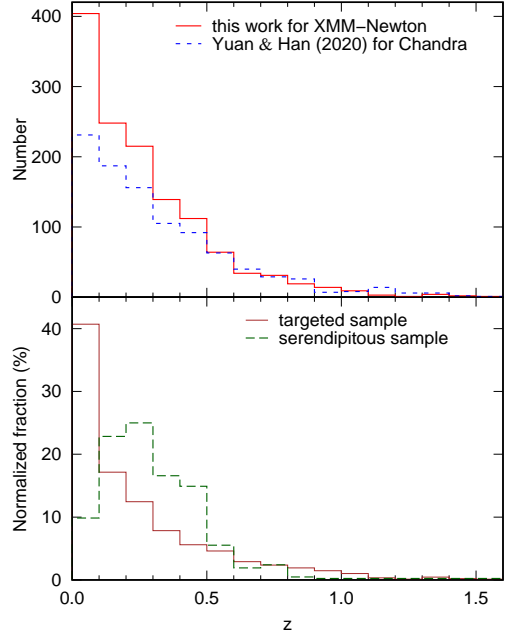}
\caption{{\it Upper panel:} Redshift distribution for 1301 of 1308
  galaxy clusters (solid) with available redshift observed by the {\it
    XMM-Newton}, compared to that (short-dashed) for clusters in
  \citet{yh20} by the {\it Chandra}. {\it Lower panel:} Redshift
  distribution for clusters in the targeted sample (solid) and the
  serendipitous sample (dashed) detected by the {\it XMM-Newton}.}
\label{redshift}
\end{figure}

Second, we combined several large cluster catalogues
\citep{aco89,pap+11,rrb+14,whl12,wh15,why18,zgx+21,kcs+21}, and check
if any clusters serendipitously observed in {\it XMM-Newton}
observations in the released dataset. The instrument mode for the MOS1
and MOS2 is set as ``{\sc full frame}'', and the modes ``{\sc full
  frame}'', ``{\sc extended full frame}'' and ``{\sc large window}''
are chosen for the EPN sytem. Observations with {\sc any} filters are
adopted. After merging multiple items in outputs, we get a sample of
416 clusters serendipitously detected in any {\it XMM-Newton}
observations. Together with 892 clusters in the targeted sample, we
get 1308 clusters in total from the {\it XMM-Newton} data achieves
(see Table \ref{tab1}).

Because of the larger field of view ($\sim30'\times30'$) of the {\it
  XMM-Newton} than that of ACIS-I ($16.9'\times16.9'$) and that of
ACIS-S ($8.3'\times50.6'$) of the {\it Chandra}, we get many galaxy
clusters at a low redshift observed by the {\it XMM-Newton}. The upper
panel of Fig. \ref{redshift} shows the redshift distribution of 1301
of 1308 clusters with available redshifts, compared with that of the
{\it Chandra} clusters, see details in \citet{yh20}.  As shown in the
lower panel of Fig.~\ref{redshift}, the targeted sample of clusters
detected by {\it XMM-Newton} are mostly of low redshifts, probably
observed for detailed studies. The serendipitous sample has larger
fraction of clusters in the middle redshift, probably as a result of
combination of a huge sample of known optical clusters and the good
{\it XMM-Newton} sensitivity.

In addition, supplementary to the {\it Chandra} sample published in
\citet{yh20}, 22 clusters are newly released by the {\it Chandra}
satellite, which are also processed in this work.

\begin{figure*}
\centering
\includegraphics[angle=270,width=0.32\textwidth]{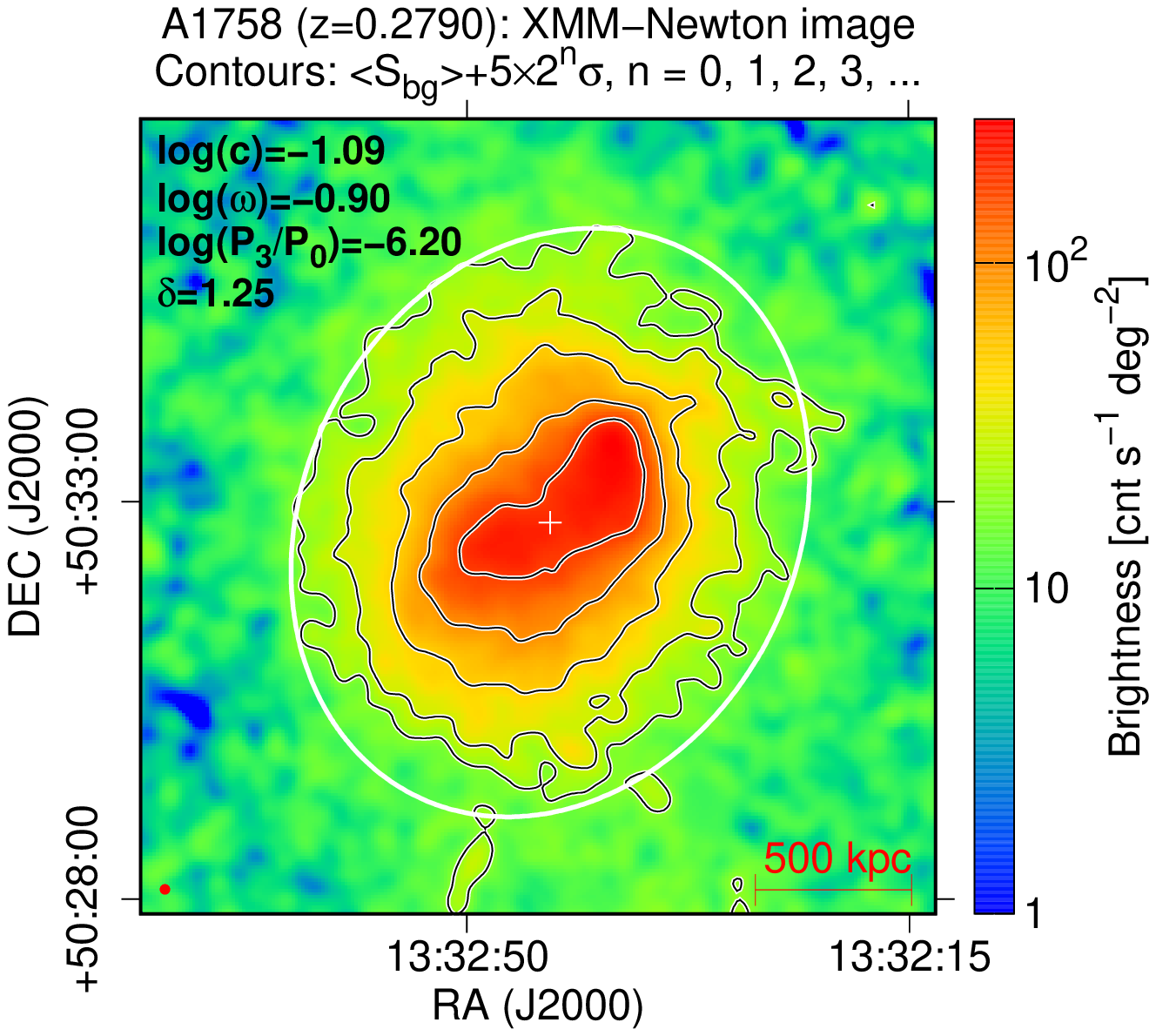}  
\includegraphics[angle=270,width=0.32\textwidth]{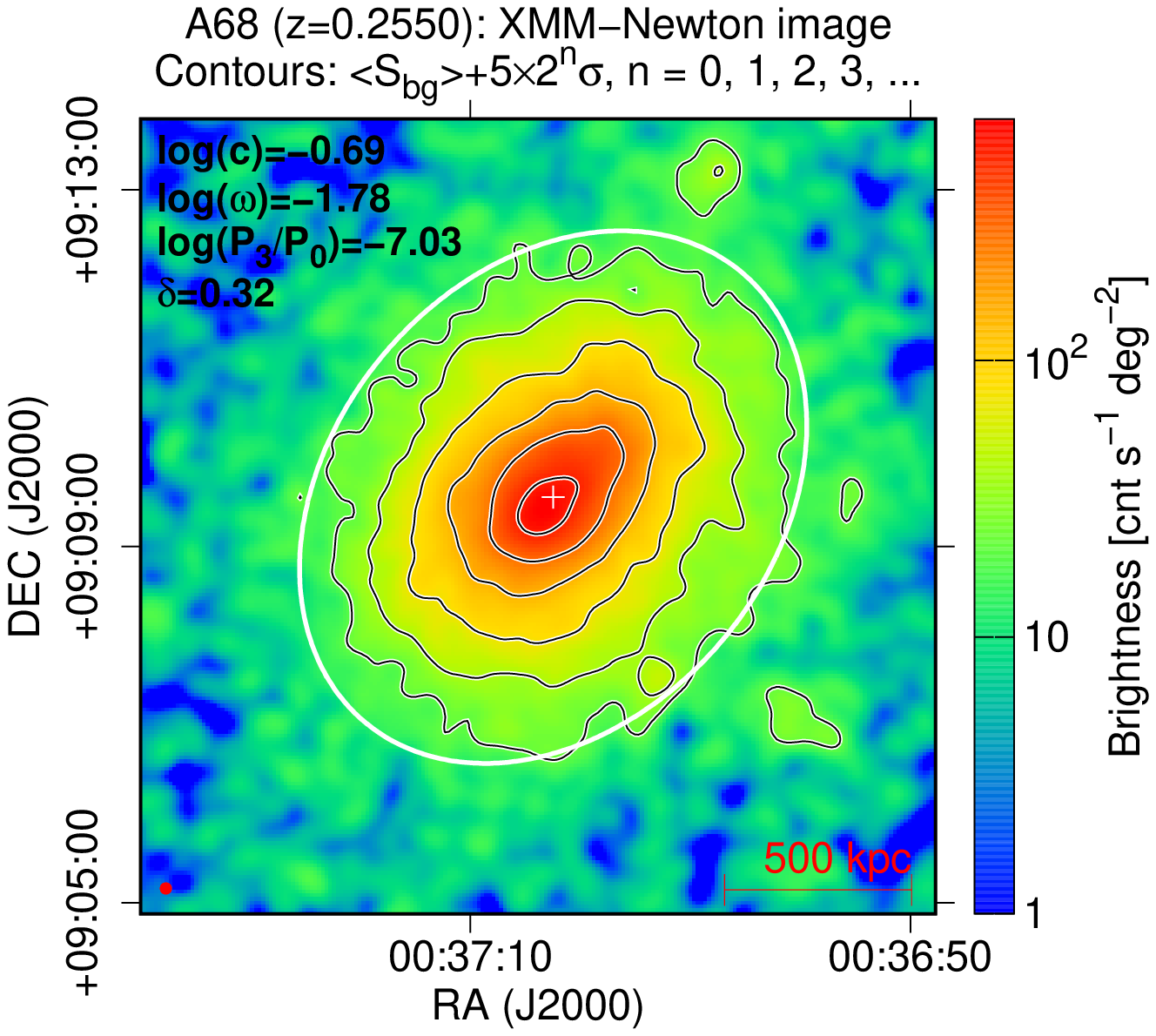} 
\includegraphics[angle=270,width=0.32\textwidth]{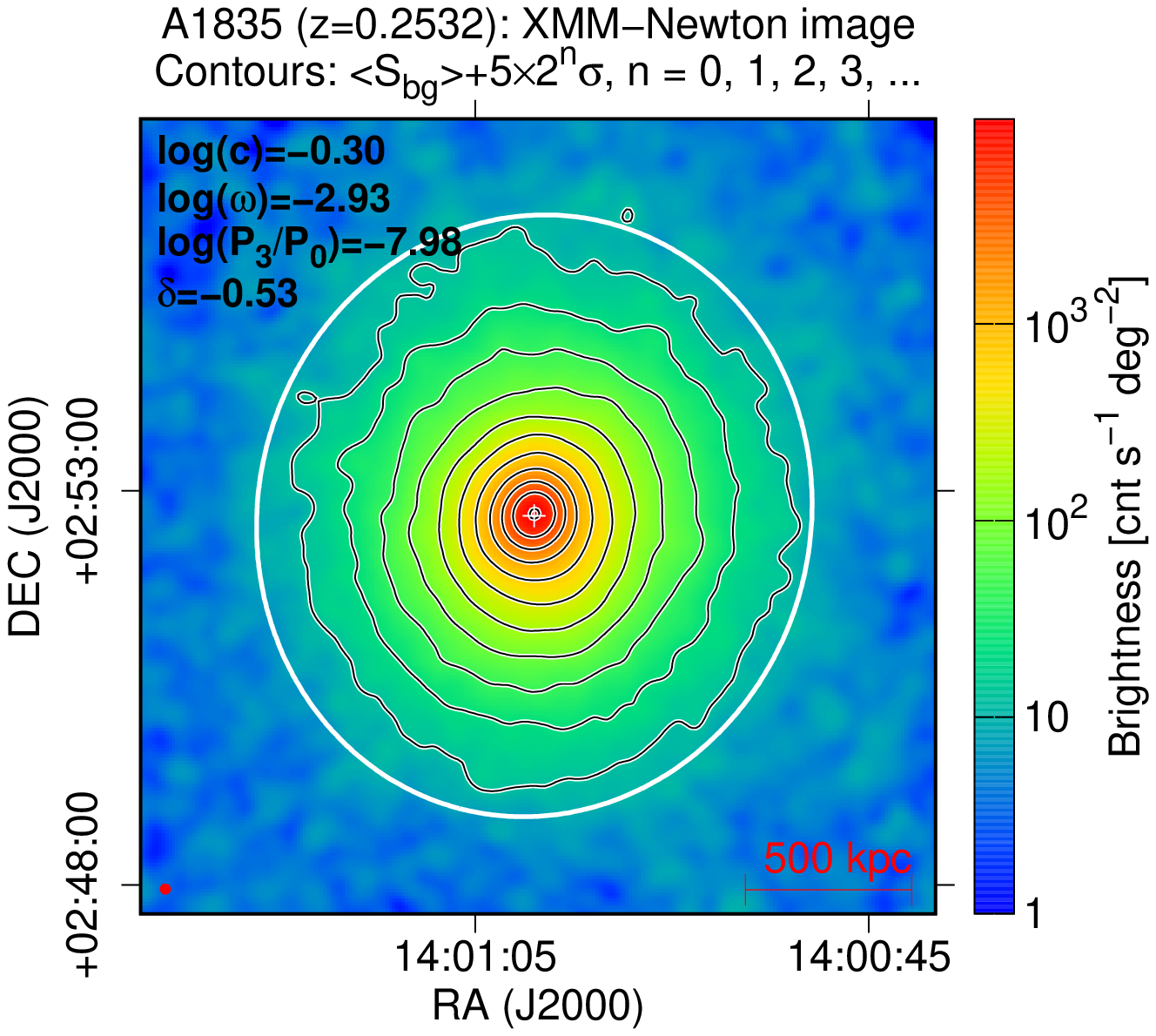}
\caption{{\it XMM-Newton} images of A1758, A68 and A1835 after data
  processing. The image fluctuation $\sigma$ is calculated in a
  surrounding clean area after the mean brightness of background
  $\langle S_{\rm bg} \rangle$ is determined. The images are fitted
  with a model as indicated by the white ellipse at the level of
  $\langle S_{\rm bg} \rangle$+5$\sigma$, with the modeled cluster
  center indicated by the white cross. The smooth scale is indicated
  as the small red circle in the bottom-left corner, and the 500 kpc
  scale is plotted on the bottom-right corner.}
\label{examples}
\end{figure*}

\begin{table*}
\setlength{\tabcolsep}{1mm}
\caption{Dynamical parameters for 1308 clusters observed by the {\it
    XMM-Newton} (see http://zmtt.bao.ac.cn/galaxy\_clusters/dyXimages/
  for the full table).}  \footnotesize
\begin{center}
  \begin{tabular}{lrrrrcccccrc}
  \hline
  \mc{1}{l}{Name}  &\mc{1}{c}{obsID} &\mc{1}{c}{RA} &\mc{1}{c}{DEC} &\mc{1}{c}{$z$}  &\mc{1}{c}{${\rm log}_{10}(c)$} &\mc{1}{c}{${\rm log}_{10}(\omega)$}  &\mc{1}{c}{${\rm log}_{10}(P_3/P_0)$} &\mc{1}{c}{$\kappa$} &\mc{1}{c}{${\rm log}_{10}(\alpha)$} &\mc{1}{c}{$\delta$} &\mc{1}{c}{Notes}\\
  \mc{1}{l}{(1)} &\mc{1}{c}{(2)} &\mc{1}{c}{(3)} &\mc{1}{c}{(4)} &\mc{1}{c}{(5)} &\mc{1}{c}{(6)} &\mc{1}{c}{(7)} &\mc{1}{c}{(8)} &\mc{1}{c}{(9)} &\mc{1}{c}{(10)} &\mc{1}{c}{(11)} &\mc{1}{c}{(12)}\\
  \hline
  RXCJ0000.1+0816    &0741581501  &0.02958 &  8.27444 &0.0396 &-0.14$\pm$0.01 &-4.04$\pm$0.08 &-6.50$\pm$0.01  &0.89 &-2.46$\pm$0.01 &-0.81$\pm$0.01 &--\\
  A2690              &0125310101  &0.09120 &-25.13822 &0.0840 &-0.60$\pm$0.01 &-1.86$\pm$0.02 &-6.66$\pm$0.05  &2.54 &-1.33$\pm$0.01 & 1.16$\pm$0.01 &--\\
  XMMXCSJ0002-3556   &0145020201  &0.56708 &-35.94272 &0.7704 &-0.64$\pm$0.03 &-1.41$\pm$0.03 &-5.81$\pm$0.11  &2.45 &-1.19$\pm$0.01 & 1.19$\pm$0.01 &--\\
  A2715              &0655300101  &0.68944 &-34.67154 &0.1160 &-0.76$\pm$0.01 &-1.52$\pm$0.01 &-5.64$\pm$0.03  &2.30 &-0.55$\pm$0.01 & 1.51$\pm$0.01 &--\\
  A2697              &0652010401  &0.79826 & -6.09169 &0.2484 &-0.66$\pm$0.02 &-2.38$\pm$0.02 &-6.67$\pm$0.13  &1.11 &-1.90$\pm$0.01 &-0.27$\pm$0.01 &--\\
  A2717              &0145020201  &0.80042 &-35.92722 &0.0490 &-0.39$\pm$0.01 &-3.32$\pm$0.01 &-7.54$\pm$0.05  &1.03 &-1.93$\pm$0.01 &-0.35$\pm$0.01 &c \\
  A2700              &0201900101  &0.96083 &  2.06333 &0.0924 &-0.51$\pm$0.01 &-3.05$\pm$0.03 &-7.88$\pm$0.10  &1.40 &-2.29$\pm$0.01 &-0.33$\pm$0.01 &--\\
  ZGXJ000402-355635  &0145020201  &1.00742 &-35.94317 &0.4974 &-0.58$\pm$0.04 &-2.12$\pm$0.05 &-6.15$\pm$0.13  &2.17 &-1.42$\pm$0.01 & 0.83$\pm$0.01 &--\\
  WHLJ000524+161309  &0783270101  &1.35000 & 16.21917 &0.1160 &-0.73$\pm$0.01 &-1.66$\pm$0.01 &-6.62$\pm$0.03  &2.05 &-1.39$\pm$0.01 & 0.76$\pm$0.01 &--\\
  A2734              &0675470801  &2.83625 &-28.85500 &0.0625 &-0.63$\pm$0.01 &-2.66$\pm$0.03 &-6.72$\pm$0.02  &1.85 &-1.37$\pm$0.01 & 0.62$\pm$0.01 &c \\
\hline
\end{tabular}
\label{tab1}
\end{center}
{Columns: (1) cluster name; (2) observation ID; (3-4) right ascension
  and declination in J2000; (5) redshift; (6) the concentration index;
  (7) the centroid shift; (8) the power ratio; (9) the profile
  parameter; (10) the asymmetry factor; (11) the morphology index;
  (12) Notes: ``--'' observed only by the {\it XMM-Newton}, ``c'' also
  observed by the {\it Chandra}.}
\end{table*}

\subsection{X-ray image processing}

The data of 1308 clusters of galaxies are processed in the {\it
  XMM-Newton} Science Analysis System \citep[SAS,][ version:
  18.0.0]{gdf+04} with the latest Current Calibration Files (CCFs)
synchronised to the CCF repository of the {\it XMM-Newton}. The
Observation Data Files (ODFs) for all clusters are downloaded with the
collected observation ID. The CCFs for a given cluster are selected
from the repository with the command {\tt cifbuild}, and the summary
file for observation information is produced by running {\tt
  odfingest}. To filter the effect of soft proton excesses, the event
files are cleaned by {\tt emchain} and {\tt mos-filter} for the MOS
systems, and by {\tt epchain} and {\tt pn-filter} for the PN
system. To be consistent with data in \citet{yh20}, only photons in
0.5-5 keV are taken. After a series of routines (e.g., {\tt cheese},
{\tt mos/pn-spectra}, {\tt mos/pn-back}, {\tt proton}, ..., see a
thread as an
example\footnote{https://www.cosmos.esa.int/web/xmm-newton/sas-thread-esasimage}),
we obtain the X-ray count image, the exposure image, the Quiescent
Particle Background (QPB) image and the soft proton image for
clusters. Images for MOS1, MOS2 and PN systems are merged with the
routine {\tt comb}. Then the background are subtracted and the
exposure are corrected by the tool {\tt adapt}. The Point Spread
Function (PSF) for each cluster is generated with the SAS tool {\tt
  psfgen}, that will be used to deconvolve the observed image.

To calculate the dynamical parameters, as done by \citet{yh20}, the
point sources in the field must be discounted. The point sources are
detected with the CIAO routine {\tt wavdetect}, and the ``holes'' of
subtracted images are filled with the tool {\tt dmfilth} according to
the ambient brightness \citep[see the Section 2.2 in][for
  details]{yh20}. To avoid possible biases caused by different
physical scales indicated by the image pixel for clusters at different
redshifts, the X-ray images of clusters are smoothed to a certain
physical scale of 30 kpc with a circular Gaussian model. For few
clusters with very low redshift $z<0.03$ the images are smoothed to 10
kpc. For 7 clusters without available redshifts, the images are
smoothed to 5 arcseconds. The processed X-ray images for all clusters
are available on the web-page:
http://zmtt.bao.ac.cn/galaxy\_clusters/dyXimages/, and three examples
are given in Fig.~\ref{examples}. The images for 22 clusters newly
released by the {\it Chandra} are processed in the same way as
\citet{yh20}.

After all images are processed, we get 351 clusters which are common
in both the {\it Chandra} and {\it XMM-Newton} databases.

\section{Parameters for dynamical states}
\label{dp}

The dynamical state of galaxy clusters can be quantitatively
determined from their X-ray images. In this section, we calculate
three widely used parameters, i.e., the concentration index $c$, the
centroid shift $\omega$ and the power ratio $P_3/P_0$, and also the
morphology index $\delta$ defined in \citet{yh20}, which is derived in
the elliptical regions adapting to the actual size and shape of
clusters. Definitions for these dynamical parameters are as
followings.

The concentration index $c$ is defined by \citet{srt+08} as being
\begin{equation}
  c=\frac{S_{\rm core}}{S_{\rm tot}}=\frac{\sum\limits_{R<\rm R_{\rm
        core}}f_{\rm obs}(x_i,y_i)}{\sum\limits_{R<\rm R_{ap}}f_{\rm
      obs}(x_i,y_i)},
\label{c}
\end{equation}
where $f_{\rm obs}(x_i,y_i)$ is the brightness corresponding to the
pixel $(x_i,y_i)$. Here the fluxes from the core and the whole regions
of clusters, $S_{\rm core}$ and $S_{\rm tot}$, are integrations of
X-ray images around the model-fitted center, i.e., the white crosses
in Fig.~\ref{examples}, rather than around the brightness peak of
images. To be consistent with previous works
\citep[e.g.,][]{ceg+10,ceb+13,yh20}, the core radius $R_{\rm core}$
and the radius for whole cluster $R_{\rm ap}$ are set as 100 kpc and
500 kpc, respectively. For relaxed clusters with a very bright core,
the gas is gradually sink to the center, so that the concentration
index $c$ should approach towards 1.0. For merging clusters with
disturbed structures, the $c$ could be very small. Because of the
large PSF of the {\it XMM-Newton} ($\sim$5'' for EPIC-PN and $\sim$6''
for EPIC-MOS, corresponding to $\sim$20 kpc if $z=0.2$), the cluster
center and the concentration index are determined from PSF-deconvolved
X-ray images \citep[e.g.,][]{lfj+17}.

For merging clusters, large offsets appear between the brightness peak
and the centroid, i.e. the fitted cluster center
\citep[e.g.,][]{kbp+01,zhk+20}. Based on the simulations by
\citet{pfb+06}, \citet{ceg+10} clearly wrote the formula as being
\begin{equation}
\omega=[\frac{1}{n-1}\sum\limits_{i}(\Delta_{i}-\langle \Delta
  \rangle)^2]^{\frac{1}{2}}\times \frac{1}{R_{\rm ap}}.
\label{w}
\end{equation}
Within the aperture radius of $R_{\rm ap} = 500$kpc, the centroid can
be found in a series of circular apertures with a decreasing step of
5\%, so that a series of offsets can be obtained. The standard
deviation of these offsets from all $i$th apertures are calculated and
normalized to $R_{\rm ap}$, which is defined as $\omega$. A small
$\omega$ implies a relaxed cluster.

Cluster mergers often trigger substructures and fluctuations in the
X-ray images \citep[see pictures in][]{me12}. \citet{bt95}
characterized substructures of clusters in various scales and defined
the ratio between $P_3$ and $P_0$ as a good dynamical proxy. The
$P_{m}$ is calculated as being 
\begin{flalign}
  \begin{split}
  &  P_{0}=[a_{0}\ln (R_{\rm ap})]^2,\\
  &  P_{m}=\frac{1}{2m^{2}R_{\rm ap}^{2m}}(a_{m}^2+b_{m}^2).
  \end{split}
\label{pm}
\end{flalign}
Often used is $m=3$ for $P_3$. The $a_{m}$ and $b_{m}$ is obtained
through
\begin{equation}
  \begin{split}
    a_{m}=\int_{r\le R_{\rm ap}}f_{\rm obs}(x_i,y_i)(r)^{m}\cos(m\phi)dx_idy_i,\\
    b_{m}=\int_{r\le R_{\rm ap}}f_{\rm obs}(x_i,y_i)(r)^{m}\sin(m\phi)dx_idy_i,
  \end{split}
\end{equation}
where $r$ is the distance of a pixel $(x_i,y_i)$ to the cluster
center, $\phi$ is the position angle of the pixel $(x_i,y_i)$.

\begin{figure*}
  \begin{center}
 \includegraphics[angle=-90,width=0.95\textwidth]{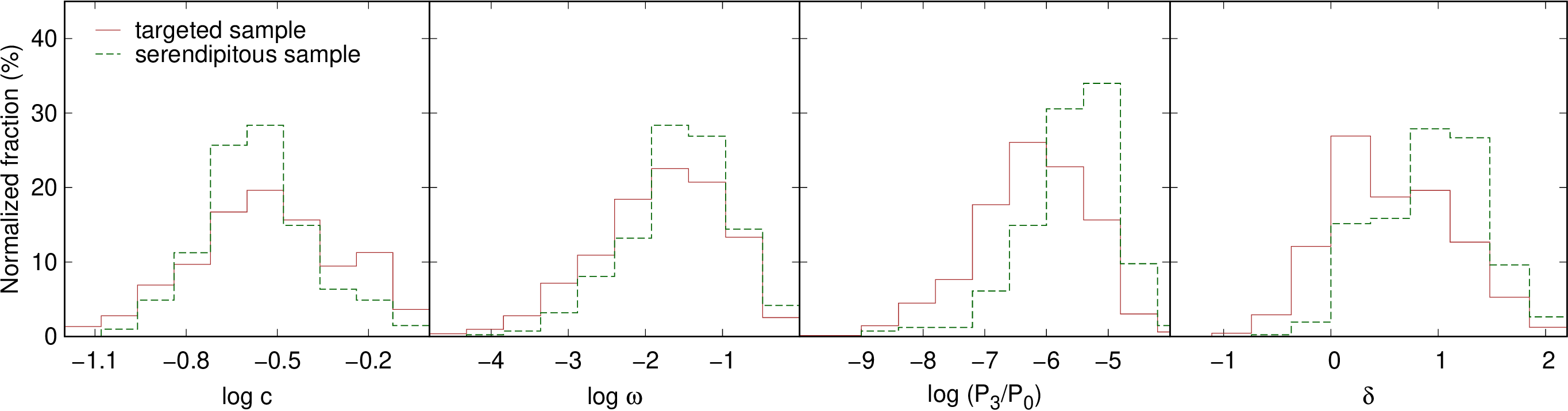}
 \caption{Distribution of dynamical parameters for clusters in the
   targeted sample (solid) and the serendipitous sample (dashed).}
\label{dyn-distribution}
\end{center}
\end{figure*}

\begin{figure*}
  \begin{center}
 \includegraphics[angle=0,width=0.7\textwidth]{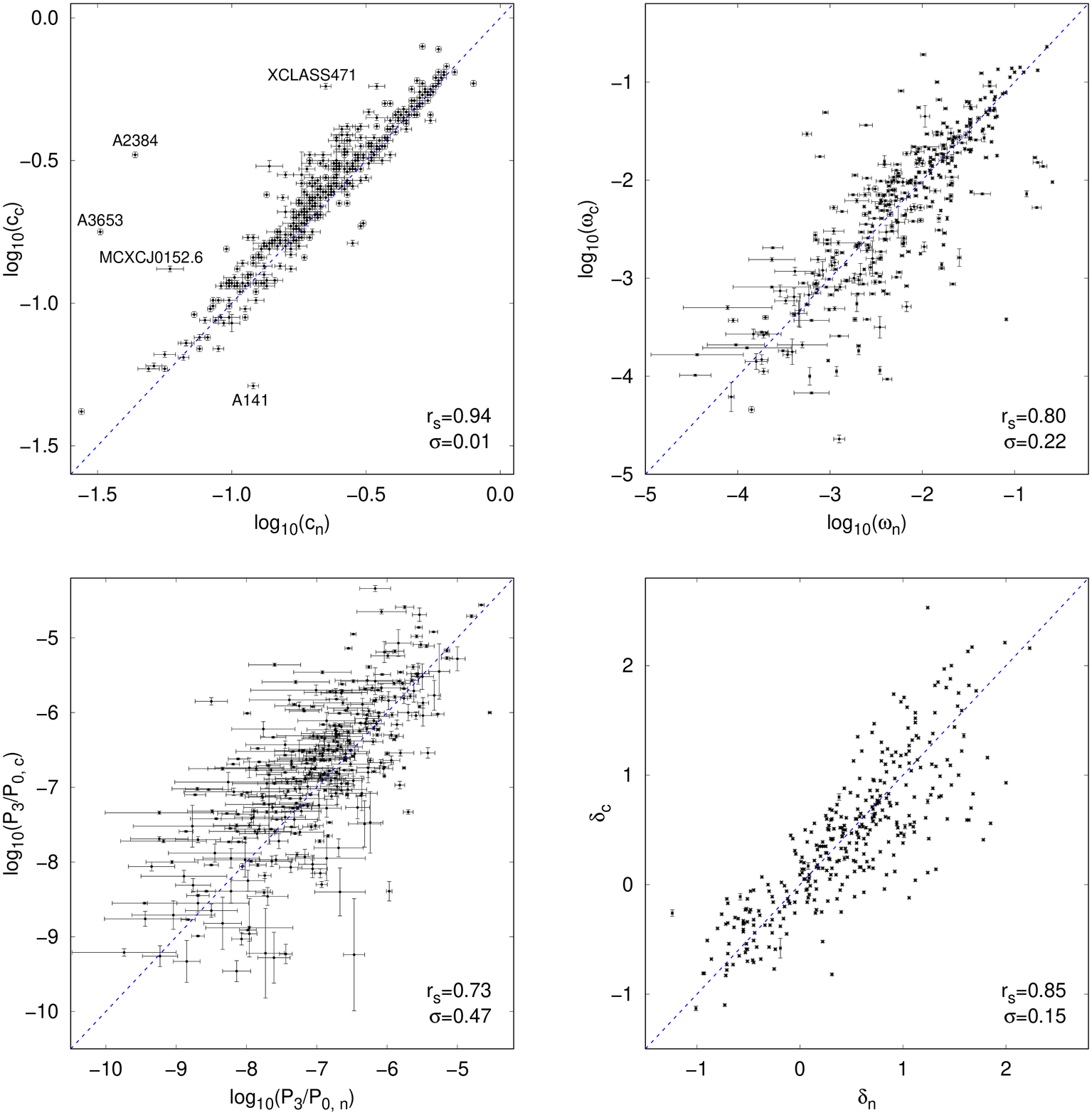}
 \caption{Comparison between the concentration indexes ({\it
     upper-left}), centroid shifts ({\it upper-right}), power ratios
   ({\it lower-left}) and morphology indexes ({\it lower-right}) for
   351 clusters derived from {\it XMM-Newton} (X-axis, with a
   subscript ``n'') and derived from {\it Chandra} images (Y-axis,
   with a subscript ``c'' ). The dotted line in each panel indicates
   equivalent values in X and Y axes. The name of 5 outliers are
   labeled in the top-left panel. The Spearman rank-order correlation
   coefficient $r_{\rm s}$ and the intrinsic scatter $\sigma$ are
   marked in each panel. }
\label{compare}
\end{center}
\end{figure*}

\begin{figure*}
\begin{center}
\includegraphics[angle=-90,width=0.99\textwidth]{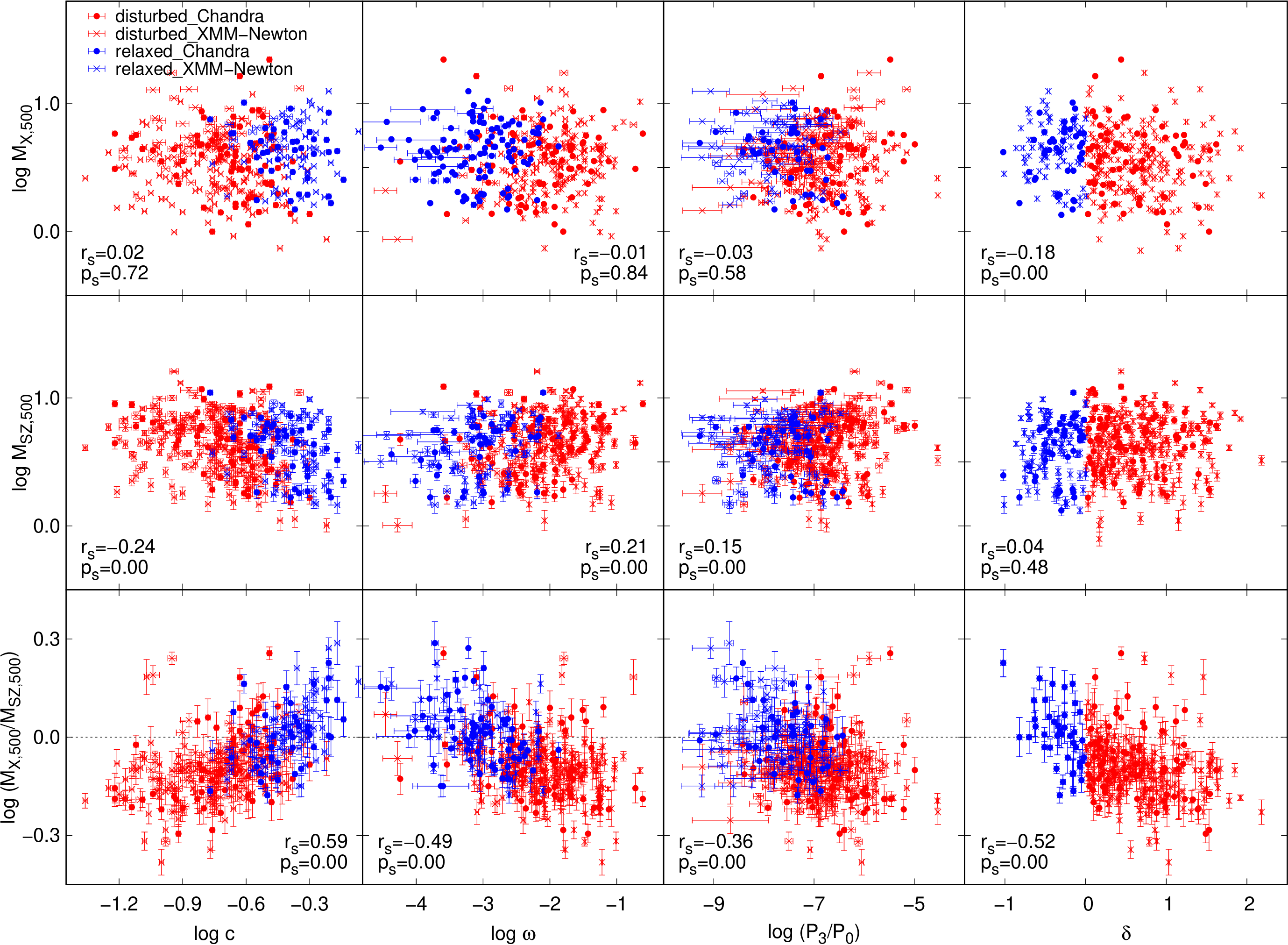}
\caption{Dependence of mass estimation on the dynamical
  parameters. {\it Top panels} are for $M_{\rm X,500}$ \citep[taken
    from][]{pap+11}, {\it middle panels} for $M_{\rm SZ, 500}$
  \citep[taken from][]{paa+16}, and {\it bottom panels} for the
  deviations of the mass estimation expressed by the mass ratio in
  logarithm. Dynamical parameters of clusters are taken from
  Table~\ref{tabA1}. Dynamical parameters derived from {\it
    XMM-Newton} images are marked as crosses, while those from {\it
    Chandra} images are presented as dots. Clusters with $\delta\ge0$
  are regarded as disturbed clusters (red), while those with
  $\delta<0$ are classified as relaxed clusters (blue). The
  correlation coefficient $r_{\rm s}$ and relevant significance
  $p_{\rm s}$ are labelled in each panel.}
\label{massdif}
\end{center}
\end{figure*}

Note that the concentration index, the centroid shift and the power
ratio are all calculated in a {\it circular} region with a given
radius of $R_{\rm ap}=500$ kpc, so that the distance or redshift of a
cluster is desired for such calculations. In principle, one can use
the mass-related radius $R_{500}$ for calculation within which the
mean density of material is 500 times of the critical density of the
Universe at the cluster redshift. \citet{yh20} found that the values
of concentration indexes, centroid shifts and power ratios derived in
500 kpc are well correlated with those from literature calculated in
$R_{500}$ or a fraction of it
\citep[e.g.,][]{wbsa13,wbc13,der+16,ajf+17}. As discussed in
\citet{yh20}, we do not use it for calculations with following
reasons. (1) The $R_{500}$ has to be derived during the process of
mass estimation, so that the observation must have sufficient X-ray
photons to derive the radial profiles of temperature and electron
density, which may not be possible for many clusters. (2) The typical
size of $R_{500}$ is about 1 Mpc \citep[e.g.,][]{pap+11}, leading a
deficient CCD coverage for clusters with $z<0.1$ (over 400 clusters,
see Fig.~\ref{redshift}). (3) The $R_{500}$ is derived under the
assumptions of spherical symmetry and virial equilibrium for
clusters. Our motivation is deriving dynamical parameters for clusters
as many as possible, here we simply take $R_{\rm ap}=500$ kpc for
calculations.

Considering the actual size and shape of clusters, \citet{yh20}
defined a morphology index $\delta$, which is adoptive to any cluster
regardless the redshift is available or not and what shape it has
(i.e. not necessary {\it circular}). First the brightness profile of
clusters is fitted with an elliptical $\beta$-model \citep{cf76}:
\begin{equation}
f_{\rm mod}(x_i,y_i)=f_{\rm mod}(r)=A(1+(\frac{r}{r_0})^2)^{-\beta}+C,
\label{Ir}
\end{equation}
where
\begin{equation}
r(x_i,y_i)=\frac{\sqrt{x^2(1-\epsilon)^2+y^2}}{1-\epsilon},
\label{r}
\end{equation}
and
\begin{equation}
  \begin{split}
    x=(x_i-x_0)\cos\theta+(y_i-y_0)\sin\theta,\\
    y=(y_i-y_0)\cos\theta-(x_i-x_0)\sin\theta.
    \label{xy}
\end{split}
\end{equation}
Here $(x_0,y_0)$, $A$, $r_0$ and $\beta$ are, respectively, the
center, the maximum, the characteristic radius and the slope index of
the best fitted model, $C$ stands for the average brightness of
background, and $\epsilon$ and $\theta$ are the fitted ellipticity and
the orientation angle. The profile parameter $\kappa$ is thus defined
as
\begin{equation}
\kappa=\frac{1+\epsilon}{\beta}.
\label{kappa}
\end{equation}
In the other hand, the asymmetry factor $\alpha$ can be calculated as being
\begin{equation}
\alpha=\frac{\sum\limits_{x_i,y_i}{[f_{\rm
        obs}(x_i,y_i)-f_{\rm
        obs}(x_i',y_i')]^2}}{\sum\limits_{x_i,y_i}{f^2_{\rm
      obs}(x_i,y_i)}} \times100\;  \%,
\label{alpha}
\end{equation}
where $(x_i,y_i)$ and $(x_i',y_i')$ are the symmetric pixel pair to
the center $(x_0,y_0)$. The morphology index $\delta$ finally defined
as the combination of the profile parameter $\kappa$ and the asymmetry
factor $\alpha$ as
\begin{equation}
\begin{aligned}
\delta &=0.68\rm{log_{10}}(\alpha)+0.73\kappa+0.21, 
\label{delta}
\end{aligned} 
\end{equation}
which can best separate the relaxed and disturbed clusters in the
$\kappa-\alpha$ space \citep{yh20}.

We calculate the concentration index $c$, the centroid shift $\omega$
and the power ratio $P_3/P_0$ in a circular region for 1308 clusters
found in the {\it XMM-Newton} archive, as listed in
Table~\ref{tab1}. The two morphology parameters $\kappa$ and $\alpha$,
and the morphology index $\delta$ are calculated in an elliptical
region with the size and shape adapting to the actual morphology of a
cluster (see the white ellipses in Fig.~\ref{examples}).

\begin{table*}
\setlength{\tabcolsep}{1mm}
\caption{Dynamical parameters for 42 clusters in \citet{yh20}
  with newly available redshifts and 22 clusters in the newly released
  {\it Chandra} data. }
\label{tab2}
\footnotesize
\begin{center}
  \begin{tabular}{lcrrcccccccc}
  \hline
  \mc{1}{l}{Name} &\mc{1}{c}{obsID} &\mc{1}{c}{RA} &\mc{1}{c}{DEC} &\mc{1}{c}{$z$}  &\mc{1}{c}{${\rm log}_{10}(c)$} &\mc{1}{c}{${\rm log}_{10}(\omega)$}  &\mc{1}{c}{${\rm log}_{10}(P_3/P_0)$} &\mc{1}{c}{$\kappa$} &\mc{1}{c}{${\rm log}_{10}(\alpha)$} &\mc{1}{c}{$\delta$}  &\mc{1}{c}{Notes}\\
  \mc{1}{l}{(1)} &\mc{1}{c}{(2)} &\mc{1}{c}{(3)} &\mc{1}{c}{(4)} &\mc{1}{c}{(5)} &\mc{1}{c}{(6)} &\mc{1}{c}{(7)} &\mc{1}{c}{(8)} &\mc{1}{c}{(9)} &\mc{1}{c}{(10)} &\mc{1}{c}{(11)} &\mc{1}{c}{(12)}\\
  \hline
 SPTCLJ0013-4906      &13462  &  3.33083 &-49.11611 &0.4075 &-0.72$\pm$0.01 &-1.50$\pm$0.01 &-5.81$\pm$0.02 &2.03 &-0.48$\pm$0.01 & 1.37$\pm$0.01 &1\\[-0.1mm]
  SPTCLJ0033-6326      &13483  &  8.46875 &-63.44389 &0.5963 &-0.56$\pm$0.01 &-2.04$\pm$0.02 &-6.05$\pm$0.07 &1.13 &-0.98$\pm$0.01 & 0.37$\pm$0.01 &1\\[-0.1mm]
  SPTCLJ0058-6145      &13479  & 14.58583 &-61.76805 &0.8300 &-0.74$\pm$0.01 &-2.20$\pm$0.01 &-5.26$\pm$0.04 &2.27 &-0.84$\pm$0.01 & 1.30$\pm$0.01 &1\\[-0.1mm]
  SPTCLJ0102-4603      &13485  & 15.67708 &-46.07195 &0.8405 &-0.82$\pm$0.01 &-1.35$\pm$0.01 &-5.56$\pm$0.11 &2.14 &-0.50$\pm$0.01 & 1.43$\pm$0.01 &1\\[-0.1mm]
  SPTCLJ0123-4821      &13491  & 20.79583 &-48.35806 &0.6550 &-0.79$\pm$0.02 &-1.82$\pm$0.01 &-6.13$\pm$0.10 &1.74 &-0.92$\pm$0.01 & 0.86$\pm$0.01 &1\\[-0.1mm]
  RXCJ0138.0-2155      &17186  & 24.51583 &-21.92528 &0.3380 &-0.29$\pm$0.01 &-3.51$\pm$0.01 &-7.91$\pm$0.15 &0.65 &-1.63$\pm$0.01 &-0.42$\pm$0.01 &1\\[-0.1mm]
  SPTCLJ0142-5032      &13467  & 25.54583 &-50.54000 &0.6793 &-0.80$\pm$0.05 &-2.01$\pm$0.06 &-5.94$\pm$0.17 &2.28 &-0.70$\pm$0.01 & 1.39$\pm$0.01 &1\\[-0.1mm]
  SPTCLJ0151-5954      &13480  & 27.85708 &-59.90805 &1.0300 &-0.86$\pm$0.01 &-1.03$\pm$0.01 &-6.57$\pm$0.16 &2.86 &-0.31$\pm$0.01 & 2.09$\pm$0.01 &1\\[-0.1mm]
  SPTCLJ0156-5541      &13489  & 29.04208 &-55.69806 &1.2200 &-0.63$\pm$0.01 &-1.72$\pm$0.02 &-7.92$\pm$0.55 &1.46 &-1.01$\pm$0.01 & 0.59$\pm$0.01 &1\\[-0.1mm]
  SPTCLJ0212-4656      &13464  & 33.10792 &-46.95000 &0.6535 &-0.90$\pm$0.01 &-1.02$\pm$0.02 &-5.32$\pm$0.04 &2.51 &-0.21$\pm$0.01 & 1.90$\pm$0.01 &1\\[-0.1mm]
  XCLASS1385           &20783  & 35.78400 & 86.33700 &0.1860 &-0.52$\pm$0.01 &-2.18$\pm$0.01 &-5.04$\pm$0.01 &2.16 &-0.59$\pm$0.01 & 1.39$\pm$0.01 &2\\[-0.1mm]
  RXJ0236.0-5224       &12096  & 39.02083 &-52.41889 &0.5900 &-0.48$\pm$0.01 &-1.93$\pm$0.02 &-5.23$\pm$0.11 &0.73 &-0.75$\pm$0.01 & 0.23$\pm$0.01 &1\\[-0.1mm]
  SPTCLJ0252-4824      &13494  & 43.21208 &-48.41500 &0.4207 &-0.92$\pm$0.01 &-1.11$\pm$0.01 &-6.33$\pm$0.06 &1.45 &-0.71$\pm$0.01 & 0.78$\pm$0.01 &1\\[-0.1mm]
  SPTCLJ0307-5042      &13476  & 46.96083 &-50.70500 &0.5500 &-0.70$\pm$0.01 &-2.76$\pm$0.02 &-7.01$\pm$0.09 &1.76 &-0.99$\pm$0.01 & 0.82$\pm$0.01 &1\\[-0.1mm]
  CXO031027-203537     &14236  & 47.61500 &-20.59306 &0.6916 &-0.47$\pm$0.01 &-2.17$\pm$0.01 &-5.97$\pm$0.03 &0.97 &-1.38$\pm$0.01 &-0.02$\pm$0.01 &1\\[-0.1mm]
  SPTCLJ0329-2330      &18282  & 52.31750 &-23.50306 &1.2270 &-0.72$\pm$0.01 &-1.97$\pm$0.03 &-6.47$\pm$0.06 &1.69 &-1.07$\pm$0.01 & 0.72$\pm$0.01 &1\\[-0.1mm]
  SPTCLJ0334-4659      &13470  & 53.54708 &-46.99611 &0.4861 &-0.44$\pm$0.01 &-2.56$\pm$0.01 &-6.86$\pm$0.07 &1.13 &-1.39$\pm$0.01 & 0.08$\pm$0.01 &1\\[-0.1mm]
  1WGAJ0337.8-2304     & 7991  & 54.46042 &-23.07278 &0.6200 &-0.43$\pm$0.02 &-2.16$\pm$0.02 &-5.84$\pm$0.12 &1.50 &-1.31$\pm$0.01 & 0.42$\pm$0.01 &1\\[-0.1mm]
  PSZG208.57-44.31     &18773  & 60.64958 &-15.67500 &0.8500 &-0.56$\pm$0.01 &-1.04$\pm$0.01 &-3.55$\pm$0.03 &2.58 &-1.37$\pm$0.01 & 1.16$\pm$0.01 &1\\[-0.1mm]
  SPTCLJ0406-4804      &13477  & 61.73083 &-48.08194 &0.7355 &-0.64$\pm$0.01 &-2.09$\pm$0.02 &-5.31$\pm$0.05 &2.04 &-0.55$\pm$0.01 & 1.33$\pm$0.01 &1\\[-0.1mm]
  PSZ1G141.73+14.22    &18289  & 70.25708 & 68.23528 &0.8330 &-0.62$\pm$0.03 &-1.79$\pm$0.03 &-7.95$\pm$0.55 &1.72 &-0.78$\pm$0.01 & 0.93$\pm$0.01 &1\\[-0.1mm]
  SPTCLJ0441-4854      &13475  & 70.45084 &-48.92389 &0.8100 &-0.49$\pm$0.02 &-2.39$\pm$0.02 &-5.87$\pm$0.11 &0.87 &-0.97$\pm$0.01 & 0.19$\pm$0.01 &1\\[-0.1mm]
  3C129.1              &19965  & 72.51458 & 45.02417 &0.0210 & 0.00$\pm$0.00 & 0.00$\pm$0.00 & 0.00$\pm$0.00 &1.87 &-1.59$\pm$0.01 & 0.49$\pm$0.01 &1\\[-0.1mm]
  SPTCLJ0456-5116      &13474  & 74.11792 &-51.27806 &0.5619 &-0.71$\pm$0.01 &-1.93$\pm$0.03 &-5.99$\pm$0.03 &1.11 &-0.92$\pm$0.01 & 0.40$\pm$0.01 &1\\[-0.1mm]
  PSZ1G127.02+26.21    &18286  & 89.84666 & 86.22861 &0.6300 &-0.79$\pm$0.04 &-1.41$\pm$0.19 &-6.10$\pm$0.26 &2.52 &-0.50$\pm$0.01 & 1.71$\pm$0.01 &1\\[-0.1mm]
  PSZ1G224.01-11.14    &18293  & 97.75291 &-14.83500 &0.6200 &-0.77$\pm$0.01 &-1.74$\pm$0.02 &-6.78$\pm$0.18 &1.91 &-0.70$\pm$0.01 & 1.13$\pm$0.01 &1\\[-0.1mm]
  SPTCLJ0655-5234      &13486  &103.97375 &-52.56805 &0.4724 &-0.76$\pm$0.01 &-1.29$\pm$0.01 &-5.82$\pm$0.01 &1.20 &-0.98$\pm$0.01 & 0.42$\pm$0.01 &1\\[-0.1mm]
  PSZ2G181.06+48.47    &22650  &144.85745 & 40.75748 &0.2350 &-0.99$\pm$0.01 &-1.10$\pm$0.01 &-5.49$\pm$0.01 &2.13 &-0.72$\pm$0.01 & 1.28$\pm$0.01 &2\\[-0.1mm]
  A1156                &21709  &166.23167 & 47.42083 &0.2152 &-0.79$\pm$0.01 &-2.40$\pm$0.01 &-8.53$\pm$0.49 &1.73 &-0.48$\pm$0.01 & 1.14$\pm$0.01 &2\\[-0.1mm]
  SDSSJ1142.8+1527     &18277  &175.69041 & 15.45778 &1.1200 &-0.52$\pm$0.02 &-2.44$\pm$0.10 &-6.23$\pm$0.12 &2.18 &-0.58$\pm$0.01 & 1.40$\pm$0.01 &1\\[-0.1mm]
  PSZ1G135.24+65.43    &21715  &184.77721 & 50.91310 &0.5271 &-0.97$\pm$0.02 &-1.01$\pm$0.01 &-5.83$\pm$0.20 &1.75 &-0.46$\pm$0.01 & 1.18$\pm$0.01 &2\\[-0.1mm]
  SPTCLJ1315-2806      &23214  &198.81100 &-28.10910 &1.3900 &-0.45$\pm$0.01 &-2.41$\pm$0.02 &-5.14$\pm$0.05 &1.24 &-0.88$\pm$0.01 & 0.52$\pm$0.01 &2\\[-0.1mm]
  WHLJ132419.7+041907  &23068  &201.08197 &  4.31863 &0.2615 &-0.26$\pm$0.01 &-3.27$\pm$0.01 &-8.84$\pm$0.12 &0.75 &-2.03$\pm$0.01 &-0.62$\pm$0.01 &2\\[-0.1mm]
  WHLJ133029.5+490848  &21708  &202.62276 & 49.14661 &0.3353 &-0.93$\pm$0.01 &-1.29$\pm$0.01 &-6.26$\pm$0.14 &2.83 &-0.48$\pm$0.01 & 1.95$\pm$0.01 &2\\[-0.1mm]
  SPTCLJ1342-2442      &21567  &205.50349 &-24.70510 &0.8100 &-0.56$\pm$0.01 &-1.52$\pm$0.01 &-5.53$\pm$0.03 &1.38 &-0.71$\pm$0.01 & 0.74$\pm$0.01 &2\\[-0.1mm]
  WHLJ134737.4+501109  &21714  &206.90582 & 50.18583 &0.2810 &-0.31$\pm$0.01 &-3.00$\pm$0.06 &-6.93$\pm$0.05 &0.85 &-1.18$\pm$0.01 & 0.03$\pm$0.01 &2\\[-0.1mm]
  A1904                &21712  &215.53278 & 48.55620 &0.0708 &-0.89$\pm$0.01 &-1.04$\pm$0.01 &-5.57$\pm$0.01 &2.30 &-0.53$\pm$0.01 & 1.52$\pm$0.01 &2\\[-0.1mm]
  PSZ2G104.74+40.42    &22641  &236.59300 & 69.95050 &0.6900 &-1.07$\pm$0.03 &-0.84$\pm$0.06 &-5.79$\pm$0.43 &1.45 &-0.26$\pm$0.01 & 1.09$\pm$0.01 &2\\[-0.1mm]
  CIZAJ1809.0-0414     &25087  &272.25500 & -4.24389 &0.3050 &-0.54$\pm$0.01 &-3.07$\pm$0.02 &-7.64$\pm$0.08 &0.94 &-1.77$\pm$0.01 &-0.31$\pm$0.01 &2\\[-0.1mm]
  PSZ2G075.85+15.53    &21544  &287.78854 & 44.91729 & ---   &      ---      &      ---      &      ---      &0.81 &-0.87$\pm$0.01 & 0.21$\pm$0.01 &2\\[-0.1mm]
  A3674                &23363  &305.20792 &-30.03361 &0.2073 &-0.48$\pm$0.01 &-2.01$\pm$0.01 &-5.73$\pm$0.02 &1.22 &-1.16$\pm$0.01 & 0.31$\pm$0.01 &2\\[-0.1mm]
  SPTCLJ2215-3537      &24614  &333.76651 &-35.62080 &1.1600 &-0.26$\pm$0.01 &-2.59$\pm$0.01 &-6.65$\pm$0.06 &0.67 &-0.89$\pm$0.01 & 0.10$\pm$0.01 &2\\[-0.1mm]
  SPTCLJ2218-4519      &13501  &334.74582 &-45.31611 &0.6365 &-0.76$\pm$0.01 &-1.68$\pm$0.01 &-5.92$\pm$0.04 &1.68 &-0.86$\pm$0.01 & 0.85$\pm$0.01 &1\\[-0.1mm]
  PSZ1G048.22-51.60    &18283  &335.05209 &-12.21639 &0.5440 &-0.72$\pm$0.01 &-1.73$\pm$0.01 &-6.21$\pm$0.05 &1.49 &-1.01$\pm$0.01 & 0.61$\pm$0.01 &1\\[-0.1mm]
  SPTCLJ2222-4834      &13497  &335.71210 &-48.57695 &0.6519 &-0.51$\pm$0.01 &-2.92$\pm$0.02 &-6.39$\pm$0.04 &1.45 &-1.22$\pm$0.01 & 0.44$\pm$0.01 &1\\[-0.1mm]
  SPTCLJ2232-6000      &13502  &338.14084 &-59.99805 &0.5948 &-0.41$\pm$0.01 &-2.89$\pm$0.01 &-6.58$\pm$0.03 &0.99 &-1.39$\pm$0.01 &-0.01$\pm$0.01 &1\\[-0.1mm]
  SPTCLJ2233-5339      &13504  &338.31876 &-53.65389 &0.4396 &-0.64$\pm$0.02 &-1.57$\pm$0.02 &-6.64$\pm$0.16 &1.10 &-1.11$\pm$0.01 & 0.26$\pm$0.01 &1\\[-0.1mm]
  SPTCLJ2236-4555      &13507  &339.21875 &-45.93000 &1.1600 &-0.48$\pm$0.01 &-2.22$\pm$0.01 &-5.31$\pm$0.07 &1.41 &-0.53$\pm$0.01 & 0.88$\pm$0.01 &1\\[-0.1mm]
  SPTCLJ2245-6207      &13499  &341.26001 &-62.11611 &0.5800 &-1.11$\pm$0.01 &-1.74$\pm$0.02 &-5.38$\pm$0.12 &2.84 &-0.22$\pm$0.01 & 2.13$\pm$0.01 &1\\[-0.1mm]
  SPTCLJ2258-4044      &13495  &344.70584 &-40.74000 &0.8971 &-0.77$\pm$0.01 &-1.92$\pm$0.01 &-6.01$\pm$0.15 &1.92 &-0.64$\pm$0.01 & 1.17$\pm$0.01 &1\\[-0.1mm]
  SPTCLJ2259-6057      &13498  &344.75208 &-60.96000 &0.7500 &-0.51$\pm$0.01 &-2.43$\pm$0.01 &-6.16$\pm$0.04 &0.99 &-1.13$\pm$0.01 & 0.16$\pm$0.01 &1\\[-0.1mm]
  SPTCLJ2301-4023      &13505  &345.47083 &-40.38889 &0.8349 &-0.41$\pm$0.01 &-1.83$\pm$0.05 &-6.00$\pm$0.05 &1.07 &-0.91$\pm$0.01 & 0.38$\pm$0.01 &1\\[-0.1mm]
  SPTCLJ2306-5120      &21549  &346.61209 &-51.33700 &1.2700 &-0.60$\pm$0.01 &-1.62$\pm$0.02 &-4.51$\pm$0.04 &1.89 &-0.79$\pm$0.01 & 1.05$\pm$0.01 &2\\[-0.1mm]
  SPTCLJ2315-2127      &21571  &348.82800 &-21.45700 &0.5400 &-0.66$\pm$0.01 &-2.45$\pm$0.03 &-5.62$\pm$0.04 &1.52 &-0.97$\pm$0.01 & 0.66$\pm$0.01 &2\\[-0.1mm]
  SPTCLJ2317-3239      &22654  &349.45499 &-32.66560 &1.0500 &-0.36$\pm$0.01 &-2.78$\pm$0.01 &-7.10$\pm$0.09 &0.90 &-1.38$\pm$0.01 &-0.07$\pm$0.01 &2\\[-0.1mm]
  SPTCLJ2320-5233      &21552  &350.12509 &-52.56410 &0.7550 &-0.48$\pm$0.04 &-2.48$\pm$0.22 &-3.39$\pm$1.28 &2.24 &-0.54$\pm$0.01 & 1.48$\pm$0.01 &2\\[-0.1mm]
  PSZ1G108.18-11.53    &17213  &350.62375 & 48.77500 &0.3347 &-1.05$\pm$0.01 &-1.52$\pm$0.01 &-5.52$\pm$0.13 &2.69 &-0.64$\pm$0.01 & 1.74$\pm$0.01 &1\\[-0.1mm]
  SPTCLJ2331-5736      &22073  &352.91891 &-57.61550 &1.3800 &-1.01$\pm$0.02 &-0.26$\pm$0.19 &-6.28$\pm$0.29 &2.87 &-0.08$\pm$0.01 & 2.25$\pm$0.01 &1\\[-0.1mm]
  SPTCLJ2334-5308      &21705  &353.51501 &-53.14100 &1.2000 &-0.39$\pm$0.03 &-2.49$\pm$0.05 &-3.98$\pm$0.07 &1.07 &-0.78$\pm$0.01 & 0.46$\pm$0.01 &2\\[-0.1mm]
  SPTCLJ2335-5434      &23027  &353.88101 &-54.58600 &0.8660 &-0.32$\pm$0.01 &-2.98$\pm$0.09 &-4.39$\pm$0.06 &1.00 &-0.90$\pm$0.01 & 0.33$\pm$0.01 &2\\[-0.1mm]
  SPTCLJ2336-5252      &23127  &354.08771 &-52.87250 &1.2200 &-0.43$\pm$0.02 &-2.08$\pm$0.08 &-3.34$\pm$0.03 &1.38 &-0.61$\pm$0.01 & 0.80$\pm$0.01 &2\\[-0.1mm]
  SPTCLJ2345-6406      &13500  &356.25000 &-64.09889 &0.9400 &-0.79$\pm$0.01 &-1.16$\pm$0.01 &-5.53$\pm$0.02 &1.42 &-0.75$\pm$0.01 & 0.74$\pm$0.01 &1\\[-0.1mm]
  SPTCLJ2352-4657      &13506  &358.06790 &-46.96000 &0.7300 &-0.66$\pm$0.01 &-2.78$\pm$0.01 &-5.68$\pm$0.05 &2.10 &-0.94$\pm$0.01 & 1.10$\pm$0.01 &1\\[-0.1mm]
  CODEX-59603          &18636  &359.43332 &  0.80194 &0.0616 &      ---      &      ---      &      ---      &0.59 &-1.69$\pm$0.01 &-0.51$\pm$0.01 &1\\[-0.1mm]
\hline
\end{tabular}
\label{tab2}
\end{center}
{Columns: (1)-(11) same as in Table 1; (12) Notes: ``1'' for  
  42 clusters with newly available redshifts; ``2'' for clusters in newly
  released data.}
\end{table*}

In addition to the 1308 clusters from the {\it XMM-Newton} archive,
the newly available redshifts are found for 42 clusters in the {\it
  Chandra} sample in \citet{yh20}, and we update their dynamical
parameters. We smooth their images to 30 kpc, and re-calculate the
dynamical parameters for these clusters, as listed in Table
\ref{tab2}. Moreover, 22 newly-archived clusters are found in the {\it
  Chandra} database, their dynamical parameters are also calculated
accordingly and listed in Table \ref{tab2}.

We present in Fig.~\ref{dyn-distribution} the distribution of
dynamical parameters for 1308 clusters in the {\it XMM-Newton} sample,
and find a continuous distribution from the disturbed to relaxed
clusters. In general, the clusters in the serendipitous sample have a
slightly higher fraction of disturbed clusters. In Fig.~\ref{compare},
we compare the values of dynamical parameters calculated based on {\it
  XMM-Newton} images and those based on {\it Chandra} images for 351
clusters which appear in both samples. They are very consistent and
the dynamical parameters concentrate around the equivalent lines.  The
Spearman rank-order correlation coefficient $r_{\rm s}$ \citep[defined
  in][p. 640]{ptvf92} indicates a great degree of correlation. The
intrinsic scatter $\sigma$ is calculated for the intrinsic dispersion
of the correlation with data uncertainties subtracted \citep[see the
  appendix of][for calculation]{yhw15}. One can see that the
concentration indexes have the best correlation among the four
dynamical parameters calculated from X-ray images of the two
satellites, except for A2384, A3653 and MCXCJ0152.6-1358, because the
cluster center is now re-defined at a position between subclusters
rather than at one subcluster by \citet{yh20}. The large departure for
XCLASS471 is mainly caused by the poor quality of image observed by
the {\it Chandra}. Also a correction was made to A141 because the
bright central part should be kept as the cluster core, but that was
mistakenly subtracted as a point source in \citet{yh20}.

In Appendix, we compose the samples of clusters, including the 964
clusters in \citet{yh20}, 22 clusters newly archived by the {\it
  Chandra}, and 1308 clusters here from {\it XMM-Newton} data. The
better estimated dynamical parameters are taken if a cluster is in
both datasets, finally the unified estimates for dynamical parameters
are listed for 1844 clusters in Table~\ref{tabA1}.

\begin{figure*}
\begin{center}
\includegraphics[angle=-90,width=0.98\textwidth]{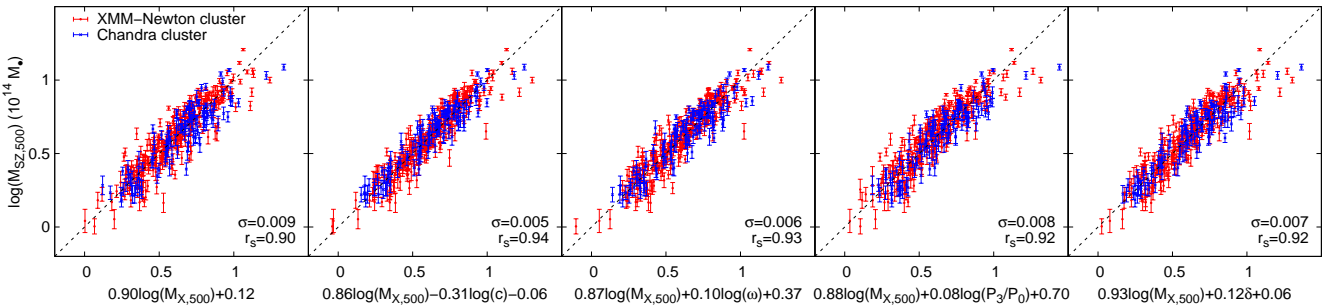}\\
\caption{Including the dynamical parameters can improve the
  consistence between the mass estimations from the X-ray luminosity
  and the SZ effect. The mass data of 388 clusters are taken from
  \citet{pap+11} and \citet{paa+16}, and the dynamical parameters from
  Table~\ref{tabA1}. The scaling relations thus can be improved as
  listed in Table~\ref{tab3}.}
\label{mass-scale}
\end{center}
\end{figure*}

\begin{table*}
  \caption{The scaling relations for mass estimations from X-ray data to that from the S-Z effect can 
    be improved by considering dynamical state. The following formula are derived from 388 clusters.}
  \label{tab3}
\begin{center}
  \begin{tabular}{lccc}
  \hline
  \mc{1}{c}{Corrections to mass scales} &\mc{1}{c}{$\sigma$} &\mc{1}{c}{$r_{\rm s}$} &\mc{1}{c}{$p_{\rm s}$}\\
  \hline
  log$_{10}(M_{\rm SZ,500})=0.90$log$_{10}(M_{\rm X,500})+0.12$                       &0.009  &0.90 &0.00\\
  log$_{10}(M_{\rm SZ,500})=0.86$log$_{10}(M_{\rm X,500})-0.31$log$_{10}(c)-0.06$      &0.005  &0.94 &0.00\\
  log$_{10}(M_{\rm SZ,500})=0.87$log$_{10}(M_{\rm X,500})+0.10$log$_{10}(\omega)+0.37$ &0.006  &0.93 &0.00\\
  log$_{10}(M_{\rm SZ,500})=0.88$log$_{10}(M_{\rm X,500})+0.08$log$_{10}(P_3/P_0)+0.70$ &0.008  &0.92 &0.00\\
  log$_{10}(M_{\rm SZ,500})=0.93$log$_{10}(M_{\rm X,500})+0.12\delta+0.06$            &0.007  &0.92 &0.00\\
 \hline
\end{tabular}
\end{center}
\end{table*}

\section{Influences of dynamical state on cluster mass estimations}
\label{appl}

Estimating masses of galaxy clusters is a key to various studies, such
as the constraint of the cosmology parameters by using clusters
\citep[e.g.,][]{vkb+09,bcc14}. For this purpose the radial profiles of
gas density and temperature of galaxy clusters should be derived first
from X-ray images for a large sample of clusters. The total masses of
clusters have been related to various observables, such as X-ray
luminosities $L_{\rm X}$, gas temperatures $T$, intracluster gas
masses $M_{\rm g}$, and the total energy of the ICM calculated through
the SZ effect $Y_{\rm SZ}$
\citep[e.g.,][]{app05,app07,m07,mem07,zfb+07,zfb+08,app+10,mae+10,rbfm11,paa+13}.
The scaling relations between masses and mass proxies are slightly
shifted for subsamples of cool-core clusters and non-cool-core
clusters \citep[e.g.,][]{crb+07,zjc+13}.

To verify the influences of dynamical state on cluster mass
estimations, we take the masses $M_{\rm X,500}$ homogeneously derived
by \citet{pap+11} for 1743 clusters mainly based on observations of
{\it ROSAT}, and the masses estimated from SZ images $M_{\rm SZ,500}$
for 1653 clusters by \citet{paa+16}. It has been widely accepted that
cluster masses estimated from the integrated SZ signal are insensitive
to the dynamical state of clusters
\citep[e.g.,][]{mhbn05,btd+07,app+10}. Among these large samples,
there are 388 clusters in our composite sample (Table~\ref{tabA1})
with dynamical parameters estimated. The X-ray and SZ mass estimates
against dynamical parameters in the top and middle panels of
Fig.~\ref{massdif} can be used to assess any dependence, the zero
value of $p_{\rm s}$ indicates a robust correlation while none-zero
value of $p_{\rm s}$ means no or fake correlation. A weak trend can be
found for the $M_{\rm X,500}-\delta$, $M_{\rm SZ,500}-c$, $M_{\rm
  SZ,500}-\omega$ and $M_{\rm SZ,500}-P_3/P_0$ correlations, though
the data scatter is rather large.

The deviation of mass estimates from the X-ray data, however, is
clearly related to dynamical parameters.  Because the X-ray mass
$M_{\rm X,500}$ is mainly derived from thermal component of the ICM,
while the SZ mass $M_{\rm SZ,500}$ is the integration of both the
thermal and non-thermal parts. The dependence on dynamical parameters
of mass estimate deviations indicate that the non-thermal energy in
the ICM plays an important role in some disturbed clusters.

Any excellent scaling relation for mass estimation should diminish
such a systematical deviation. If the SZ mass is taken as an ideal
mass unaffected by the dynamical state of clusters, the scaling
relations for the mass estimation from the X-ray measurements should
include the small but systematical deviations caused by the dynamical
state shown in the lowest panels of Fig.~\ref{massdif}. As shown in
the most left panel of Fig.~\ref{mass-scale} and also parameters in
Table~\ref{tab3}, the data scatter (i.e. the standard deviation
$\sigma$) is the largest around the best fitted scaling line. However,
when the dynamical parameters are included as the second term in the
Table~\ref{tab3}, the scatters in the other panels of
Fig.~\ref{massdif} become smaller, especially for those data with
large-deviations. The correlation parameter $r_{\rm s}$ in these plots
is also improved (closer to 1.0), as listed in Table~\ref{tab3},
though by only a small amount because of smaller weighted
contributions to the mass estimation from dynamical parameters than
that from the X-ray measurements themselves.

\section{Summary}
\label{summary}
Dynamical state can be revealed from X-ray images. Following
\citet{yh20}, we derive dynamical parameters for clusters of galaxies
archived by the {\it XMM-Newton} satellite. We got data for 892
clusters from targeted observations, and also found 416 clusters
serendipitously observed. The {\it XMM-Newton} images for these 1308
clusters are processed with the same procedures.  We smoothed the
X-ray images to a given physical scale of 30 kpc to calculate the
concentration index, the centroid shift, the power ratio. The
morphology index are calculated according to image sizes and shapes
for all these clusters.

In addition, we calculated the dynamical parameters for 42 clusters in
the {\it Chandra} sample of \citet{yh20} because of the newly
available redshifts, and also for 22 clusters in the newly released
{\it Chandra} data.

Based on the results for 351 clusters which appear in both the {\it
  XMM-Newton} sample and {\it Chandra} sample, we found that dynamical
parameters derived from {\it XMM-Newton} images are well consistent with
those calculated with {\it Chandra} images.

We found the influence of dynamical state on cluster mass
estimation. The scaling relation for masses estimation from the X-ray
can be improved by including the dynamical parameters.

\section*{Acknowledgements}

We thank the referee for instructive comments which improve this paper
significantly. The authors are supported by the National Natural
Science Foundation of China (11803046, 11988101, 12073036). YZS
acknowledges the support by the science research grants from the China
Manned Space Project (No. CMS-CSST-2021-A01, CMS-CSST-2021-B01).
This work is based on observations obtained with {\it XMM-Newton}, an
ESA science mission with instruments and contributions directly funded
by ESA Member States and NASA. This research has used data obtained
from the {\it Chandra Data Archive} and the {\it Chandra Source
  Catalogue}, and software provided by the the {\it Chandra X-ray
  Center (CXC)} in the application packages {\sc ciao}, {\sc chips},
and {\sc sherpa}. This research has made use of the NASA/IPAC
Extragalactic Database (NED), which is funded by the National
Aeronautics and Space Administration and operated by the California
Institute of Technology.

\section*{Data availability}

The data underlying this article, including the full Table~\ref{tab1}
and Table~\ref{tabA1}, the X-ray images for all clusters and the code
for calculations, are available on the web-page
http://zmtt.bao.ac.cn/galaxy\_clusters/dyXimages/.

\bibliographystyle{mnras}
\bibliography{ref}

\begin{appendix}
\section{Dynamical parameters derived from X-ray images of Chandra and XMM-Newton}
\label{appndx}

\begin{table*}
\setlength{\tabcolsep}{1mm}
\caption{Dynamical parameters for 1844 clusters of galaxies derived
  from {\it Chandra} and {\it XMM-Newton} images. The full table is
  available on http://zmtt.bao.ac.cn/galaxy\_clusters/dyXimages/.}
\label{tabA1}
\footnotesize
\begin{center}
  \begin{tabular}{lrrrcccccccc}
  \hline
  \mc{1}{l}{Name} &\mc{1}{c}{obsID} &\mc{1}{c}{RA} &\mc{1}{c}{DEC} &\mc{1}{c}{$z$}  &\mc{1}{c}{${\rm log}_{10}(c)$} &\mc{1}{c}{${\rm log}_{10}(\omega)$}  &\mc{1}{c}{${\rm log}_{10}(P_3/P_0)$} &\mc{1}{c}{$\kappa$} &\mc{1}{c}{${\rm log}_{10}(\alpha)$} &\mc{1}{c}{$\delta$}  &\mc{1}{c}{Note}\\
  \mc{1}{l}{(1)} &\mc{1}{c}{(2)} &\mc{1}{c}{(3)} &\mc{1}{c}{(4)} &\mc{1}{c}{(5)} &\mc{1}{c}{(6)} &\mc{1}{c}{(7)} &\mc{1}{c}{(8)} &\mc{1}{c}{(9)} &\mc{1}{c}{(10)} &\mc{1}{c}{(11)} &\mc{1}{c}{(12)}\\
  \hline
  RXCJ0000.1+0816      &0741581501  &0.02958 &  8.27444 &0.0396 &-0.14$\pm$0.01 &-4.04$\pm$0.08 &-6.50$\pm$0.01 &0.89 &-2.46$\pm$0.01 &-0.81$\pm$0.01 &2\\
  A2690                &0125310101  &0.09120 &-25.13822 &0.0840 &-0.60$\pm$0.01 &-1.86$\pm$0.02 &-6.66$\pm$0.05 &2.54 &-1.33$\pm$0.01 & 1.16$\pm$0.01 &2\\
  SPTCLJ0000-5748      &     18238  &0.25000 &-57.80695 &0.7020 &-0.29$\pm$0.01 &-3.50$\pm$0.01 &-8.15$\pm$0.22 &0.80 &-1.19$\pm$0.01 &-0.02$\pm$0.01 &1\\
  SPTCLJ0001-5440      &     19761  &0.40583 &-54.66972 &0.7300 &-0.50$\pm$0.01 &-1.53$\pm$0.01 &-3.99$\pm$0.05 &1.40 &-0.34$\pm$0.01 & 1.00$\pm$0.01 &1\\
  XMMXCSJ0002-3556     &0145020201  &0.56708 &-35.94272 &0.7704 &-0.64$\pm$0.03 &-1.41$\pm$0.03 &-5.81$\pm$0.11 &2.45 &-1.19$\pm$0.01 & 1.19$\pm$0.01 &2\\
  A2715                &0655300101  &0.68944 &-34.67154 &0.1160 &-0.76$\pm$0.01 &-1.52$\pm$0.01 &-5.64$\pm$0.03 &2.30 &-0.55$\pm$0.01 & 1.51$\pm$0.01 &2\\
  A2697                &0652010401  &0.79826 & -6.09169 &0.2484 &-0.66$\pm$0.02 &-2.38$\pm$0.02 &-6.67$\pm$0.13 &1.11 &-1.90$\pm$0.01 &-0.27$\pm$0.01 &2\\
  A2717                &0145020201  &0.80042 &-35.92722 &0.0490 &-0.39$\pm$0.01 &-3.32$\pm$0.01 &-7.54$\pm$0.05 &1.03 &-1.93$\pm$0.01 &-0.35$\pm$0.01 &4\\
  A2700                &0201900101  &0.96083 &  2.06333 &0.0924 &-0.51$\pm$0.01 &-3.05$\pm$0.03 &-7.88$\pm$0.10 &1.40 &-2.29$\pm$0.01 &-0.33$\pm$0.01 &2\\
  ZGXJ000402-355635    &0145020201  &1.00742 &-35.94317 &0.4974 &-0.58$\pm$0.04 &-2.12$\pm$0.05 &-6.15$\pm$0.13 &2.17 &-1.42$\pm$0.01 & 0.83$\pm$0.01 &2\\
\hline
\end{tabular}
\end{center}
{Columns: (1) cluster name; (2) observation ID; (3-4) right ascension
  and declination in J2000; (5) redshift; (6) the concentration index;
  (7) the centroid shift; (8) the power ratio; (9) the profile
  parameter; (10) the asymmetry factor; (11) the morphology index;
  (12) Note: ``1'' cluster only observed by {\it Chandra}; ``2'' only
  observed by {\it XMM-Newton}; ``3'' cluster with both {\it Chandra}
  and {\it XMM-Newton} images, parameters derived from {\it Chandra}
  images are taken; ``4'' cluster with both but parameters from {\it
    XMM-Newton}; ``5'' cluster from newly released archive of {\it
    Chandra}, ``6'' clusters from {\it Chandra} with newly available
  redshift.}
\end{table*}

To ease the usage of dynamical parameters for 964 clusters in
\citet{yh20}, for 64 clusters updated in this paper with {\it Chandra}
images, and for 1308 clusters with {\it XMM-Newton} images, we make an
uniform list for 1844 clusters, as presented in Table~\ref{tabA1}. For
clusters with both {\it Chandra} and {\it XMM-Newton} observations, we
choose the one with better quality of imaging, such as higher
signal-to-noise ratio (e.g., MCXCJ0152.6-1358 and PSZ2G243.15-73.84)
or better CCD coverage (e.g., A13 and A399).

\end{appendix}

\label{lastpage}
\end{document}